\documentclass[prd,10pt,twocolumn,showpacs,nofootinbib,aps,superscriptaddress,preprintnumbers]{revtex4-1}

\usepackage{bm}
\usepackage{graphicx}
\usepackage{multirow}
\usepackage{amsmath}
\usepackage{mathrsfs}
\usepackage{amssymb}
\usepackage{hyperref}
\usepackage{yfonts}
\usepackage{color}
\usepackage{xspace}
\usepackage{verbatim}
\usepackage{mathtools}
\usepackage{booktabs}
\usepackage{tabularx}
\usepackage[utf8x]{inputenc} 
\usepackage[dvipsnames,table]{xcolor}
\usepackage{soul}
\hypersetup{urlcolor=RedViolet,
	    citecolor=ForestGreen ,
	    linkcolor=RoyalBlue, 
	    colorlinks=true}
\definecolor{lightgray}{rgb}{0.9,0.9,0.9}	    
\definecolor{green}{rgb}{0,0.5,0}
\definecolor{red}{rgb}{1,0,0}
\definecolor{blue}{rgb}{0,0,0.5}

\newcommand{\ra}[1]{\renewcommand{\arraystretch}{#1}}
\renewcommand\({\left(}
\renewcommand\){\right)}
\renewcommand\[{\left[}
\renewcommand\]{\right]}
\newcommand{\bb}[2]{ \left\{ \begin{array}{ll} #1 \\ #2 \end{array}  \right.}
\newcommand{\be}{\begin{equation}}
\newcommand{\ee}{\end{equation}}
\newcommand{\bea}{\begin{eqnarray}}
\newcommand{\eea}{\end{eqnarray}}

\newcommand{\sinc}{\text{sinc}}
\DeclareMathAlphabet{\mathpzc}{OT1}{pzc}{m}{it}

\newcommand{\m}{m_a}
\newcommand{\gag}{g_{a\gamma}}

\newcommand{\Fa}{N_{\gamma_{10}}}

\begin{document}

\title{Weighing the Solar Axion}
%\title{Measuring the Axion Mass With IAXO}
%\title{Measuring the Axion Mass With Helioscopes}

\author{Theopisti Dafni}
\affiliation{Departamento de F\'isica Te\'orica, Universidad de
  Zaragoza, Pedro Cerbuna 12, E-50009, Zaragoza, Espa\~{n}a}
  
  \author{Ciaran A. J.~O'Hare}
  \email{ciaran.aj.ohare@gmail.com}
\affiliation{Departamento de F\'isica Te\'orica, Universidad de
  Zaragoza, Pedro Cerbuna 12, E-50009, Zaragoza, Espa\~{n}a}

\author{Biljana~Laki\' c}
\affiliation{Rudjer Bo\v skovi\' c Institute, Bijeni\v{c}ka 54, 10000 Zagreb, Croatia}

\author{Javier Gal\' an}
\affiliation{Departamento de F\'isica Te\'orica, Universidad de
  Zaragoza, Pedro Cerbuna 12, E-50009, Zaragoza, Espa\~{n}a}
  
\author{Francisco J. Iguaz}
\affiliation{Departamento de F\'isica Te\'orica, Universidad de
  Zaragoza, Pedro Cerbuna 12, E-50009, Zaragoza, Espa\~{n}a}
  \affiliation{Synchrotron Soleil, BP 48, Saint-Aubin, 91192 Gif-sur-Yvette, France}
  
\author{Igor G. Irastorza}
\affiliation{Departamento de F\'isica Te\'orica, Universidad de
  Zaragoza, Pedro Cerbuna 12, E-50009, Zaragoza, Espa\~{n}a}

\author{Kre\v{s}imir~Jakov\v ci\' c}
\affiliation{Rudjer Bo\v skovi\' c Institute, Bijeni\v{c}ka 54, 10000 Zagreb, Croatia}

\author{Gloria Luz\' on}
\affiliation{Departamento de F\'isica Te\'orica, Universidad de
  Zaragoza, Pedro Cerbuna 12, E-50009, Zaragoza, Espa\~{n}a}

\author{Javier Redondo}
\affiliation{Departamento de F\'isica Te\'orica, Universidad de
  Zaragoza, Pedro Cerbuna 12, E-50009, Zaragoza, Espa\~{n}a}
\affiliation{Max-Planck-Institut f\"ur Physik (Werner-Heisenberg-Institut), F\"ohringer Ring 6, 80805 M\"unchen, Germany}

\author{Elisa Ruiz Ch\'oliz}
\affiliation{Departamento de F\'isica Te\'orica, Universidad de
  Zaragoza, Pedro Cerbuna 12, E-50009, Zaragoza, Espa\~{n}a}

\date{\today}
\smallskip
\begin{abstract}
Axion helioscopes search for solar axions and axion-like particles via inverse Primakoff conversion in strong laboratory magnets pointed at the Sun. While helioscopes can always measure the axion coupling to photons, the conversion signal is independent of the mass for axions lighter than around 0.02 eV. Masses above this value on the other hand have suppressed signals due to axion-photon oscillations which destroy the coherence of the conversion along the magnet. However, the spectral oscillations present in the axion conversion signal between these two regimes are highly dependent on the axion mass. We show that these oscillations are observable given realistic energy resolutions and can be used to determine the axion mass to within percent-level accuracies. Using projections for the upcoming helioscope IAXO, we demonstrate that $>3\sigma$ sensitivity to a nonzero axion mass is possible between $3 \times 10^{-3}$ and $10^{-1}$ eV for both the Primakoff and axion-electron solar fluxes. 
\end{abstract}

\maketitle

\section{Introduction} 
\label{sec:intro}
Axions~\cite{Peccei:1977hh, Peccei:1977ur, Weinberg:1977ma, Wilczek:1977pj, Kim:2008hd} and axion-like particles~\cite{Jaeckel:2010ni} (hereafter referred to as axions) are low-mass pseudoscalars that are expected to couple extremely weakly to standard model fields. The coupling of axions to two photons is of particular interest as it is guaranteed---barring any accidental cancellations---for `QCD' axions involved in the well-known solution to the strong $CP$ problem~\cite{Peccei:1977hh, Peccei:1977ur}. This coupling is potentially observable if one is able to coherently boost axion-photon conversion inside a strong macroscopic magnetic field~\cite{Sikivie:1983ip, Raffelt:1987im, vanBibber:1988ge}.  Three established experimental approaches use this coherent conversion: {\it light-shining-through-wall} (LSW) experiments use high-intensity light sources and strong magnetic fields to produce axions in a laboratory~\cite{Ballou:2015cka, DellaValle:2015xxa, Bahre:2013ywa}; {\it haloscopes} search for relic axions that may constitute the dark matter (DM) halo of our galaxy~\cite{Asztalos:2010,Brubaker:2016ktl,Ouellet:2018beu,TheMADMAXWorkingGroup:2016hpc,Alesini:2017ifp,Goryachev:2017wpw,McAllister:2018ndu,Kahn:2016aff}; and {\it helioscopes} search for the axions that may be emitted by the Sun~\cite{Zioutas:2004hi, Andriamonje:2007ew, Arik:2008mq, Arik:2011rx, Arik:2013nya,Anastassopoulos:2017ftl}. See Ref.~\cite{Irastorza:2018dyq} for a recent review of experimental searches for axions.

A helioscope technique aims to observe the precious few solar axions converting into x-rays inside a long transverse magnetic field. The expected number of converted photons is given by a convolution of the solar axion spectrum and the axion-photon conversion probability. For axion masses below a critical value the conversion probability is effectively constant over the full spectrum of axions but the resulting signal is insensitive to the value of the mass, $m_a$. Above this value, the photons converting from axions at different positions along the magnet interfere destructively. Although this destructive interference reduces the strength of the observable signal overall, the resulting spectral oscillations introduce a strong dependence on $m_a$. Hence with a large enough and sensitive enough helioscope, and an x-ray detector with good energy resolution, a measurement of the axion mass may in fact be possible. This effect was highlighted as potentially exploitable for CAST in Ref.~\cite{Arik:2008mq}, but a detailed exploration of its utility for IAXO has not yet been performed. Understanding how the effect can be used is important if we wish to measure the axion mass below $\sim$0.01 eV. The feasible step sizes of a buffer gas density scan below this point would not be small enough to make a measurement that distinguishes the massive axion from $m_a = 0$.

An investigation into the measurability of the axion mass in a helioscope is extremely well motivated. Firstly, whilst the presence of a significant x-ray flux above background may well imply the existence of a new pseudoscalar behaving like an axion, only a measurement of the mass can enable a particle identification. This step is essential for relating the new particle to a QCD axion model or otherwise.
Moreover, a helioscopic measurement of the mass may well turn out to be a crucial step in discovering dark matter. The most powerful technologies for axion haloscopes require the precise tuning of a device into the frequency of galactic axion oscillations; given essentially by the axion mass. Since the signal is very weak, and technological restrictions limit the range of frequencies over which any one experiment can operate at one time, the search for DM axions usually requires very slow scans over very narrow bandwidths. Even then, whilst a positive signal does immediately provide the axion mass, the coupling to photons, $g_{a\gamma}$, will require a different measurement. In contrast to helioscopes, a haloscope has the orthogonal problem in that they are only sensitive to the combination $g_{a\gamma}^2\xi_a \rho_0$, where $\rho_0$ is the local DM density, and $\xi_a$ the fraction made up of axions. In the event of a positive signal, a haloscope cannot provide $g_{a\gamma}$, unless an assumption is made on $\xi_a$. 

A helioscope is therefore manifestly complementary. For example, if the axion is found first by IAXO, this detection would serve as an input to design DM searches. Experiments for this mass range such as the proposed optical haloscopes~\cite{Baryakhtar:2018doz} are fraught with difficulty and require fine-tuned designs with limited ranges of mass sensitivity. Nevertheless the motivation is strong, if a detection was possible in both a haloscope and a helioscope this would eventually enable the determination of the fraction of dark matter in axionic form\footnote{It is worth noting that direct searches for any DM candidate are generally completely incapable of determining how much of the local DM distribution is comprised of the detected particle.}. In the range of masses accessible to a helioscope, the QCD axion is generally associated with being a subdominant DM component in the simplest cosmologies. Although given the uncertainties, it could in fact easily account for all of it in both the postinflationary~\cite{Kawasaki:2014sqa,Gorghetto:2018myk,Ringwald:2015dsf,Co:2017mop}, and preinflationary scenarios~\cite{Visinelli:2009kt,Hoof:2018ieb}. Furthermore there are long-standing astrophysical anomalies for which an axion in this range has been shown to be a viable explanation~\cite{Isern:1992gia,Isern:2008nt,Corsico:2012ki,Corsico:2012sh,Viaux:2013lha,Corsico:2014mpa,Ayala:2014pea,Bertolami:2014noa,Arceo-Diaz:2015pva,Giannotti:2015kwo,Corsico:2016okh,Battich:2016htm}.

\begin{figure}[t]
\includegraphics[width=0.49\textwidth] {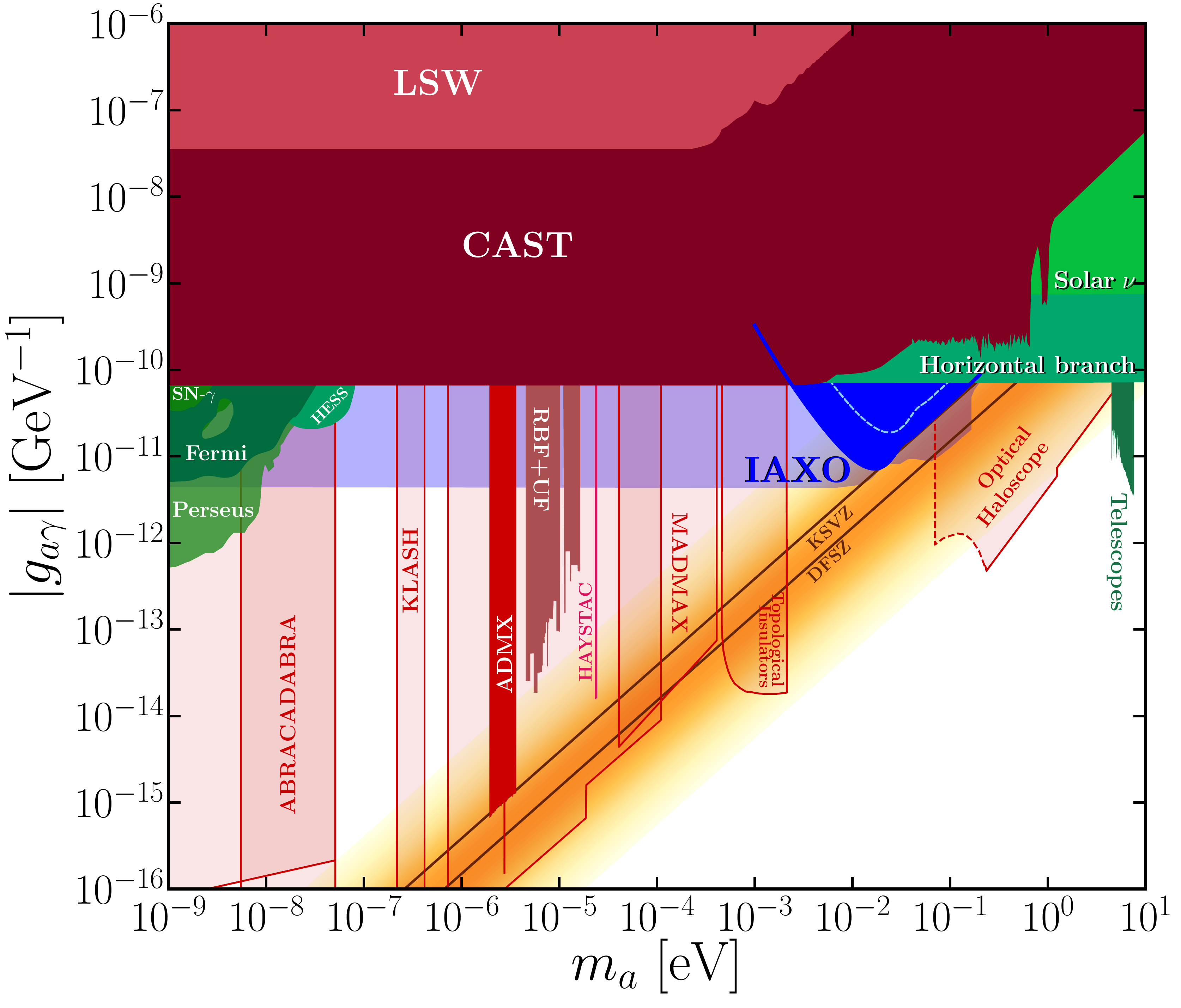}
\caption{\label{fig:mainresult} The axion parameter space $g_{a\gamma}$-$m_a$. The main result of this paper is shown as an opaque blue region: the median range of masses and couplings for which IAXO can determine the axion mass to be nonzero with 3$\sigma$ significance. Within this region we also show the limit for babyIAXO as a lighter dashed line. The QCD axion band is shaded in orange. In various shades of green are axions excluded by astrophysical arguments, and in red are experimental bounds, cited in the main text. We also show projected sensitivities as transparent regions. In particular we show the `optical haloscope' proposal of Ref~\cite{Baryakhtar:2018doz} which overlaps with our result and would hence be directly complementary.}
\end{figure}

We investigate the observability of the axion mass in the International Axion Observatory (IAXO)~\cite{Irastorza:2013dav, Armengaud:2014gea}. IAXO aims to be sensitive to QCD axion-photon couplings, down to a few $10^{-12} \, \, \rm{ GeV}^{-1}$. This value is more than one order of magnitude beyond the existing experimental and astrophysical bounds, shown in Fig.~\ref{fig:mainresult}. Along with these bounds we display the main result of this paper: the median limit for a 3$\sigma$ discovery of the axion mass in the vacuum phase of IAXO (and babyIAXO, discussed in Sec.~\ref{sec:baby}). We shade in various hues of red (green) experimental (astrophysical) exclusion limits (opaque) and projections (transparent): solar neutrinos~\cite{Vinyoles:2015aba}, horizontal branch stars~\cite{Ayala:2014pea}, SN1987A~\cite{Payez:2014xsa}, the Perseus cluster~\cite{Malyshev:2018rsh}, H.E.S.S.~\cite{Abramowski:2013oea}, Fermi-LAT~\cite{TheFermi-LAT:2016zue}, telescopes~\cite{Grin:2006aw}, LSW~\cite{Ballou:2015cka, DellaValle:2015xxa}, CAST~\cite{Anastassopoulos:2017ftl}, RBF+UF~\cite{DePanfilis:1987dk,Hagmann:1990tj}, ADMX~\cite{Du:2018uak}, HAYSTAC~\cite{Brubaker:2016ktl}, ABRACADABRA~\cite{Foster:2017hbq}, MADMAX~\cite{TheMADMAXWorkingGroup:2016hpc}, KLASH~\cite{Alesini:2017ifp}, topological insulators~\cite{Marsh:2018dlj}, and optical haloscopes~\cite{Baryakhtar:2018doz}.

The plan of the paper is as follows. In Sec.~\ref{sec:helio} we outline the calculation of an axion signal inside IAXO and describe the axion mass dependence that we use to derive projections for its discovery using the statistical methodology outlined in Sec.~\ref{sec:stats}. In Sec.~\ref{sec:results} we demonstrate how well IAXO can distinguish the massive axion from a massless axion, and the accuracy to which the axion mass can be measured. In Sec.~\ref{sec:conc} we conclude. 

The results presented in this work are reproducible via publicly available python notebooks\footnote{\url{https://cajohare.github.io/IAXOmass}}.

\section{The axion mass in a helioscope}
\label{sec:helio}
\subsection{Solar axions}

\begin{figure}[t]
\includegraphics[width=0.49\textwidth] {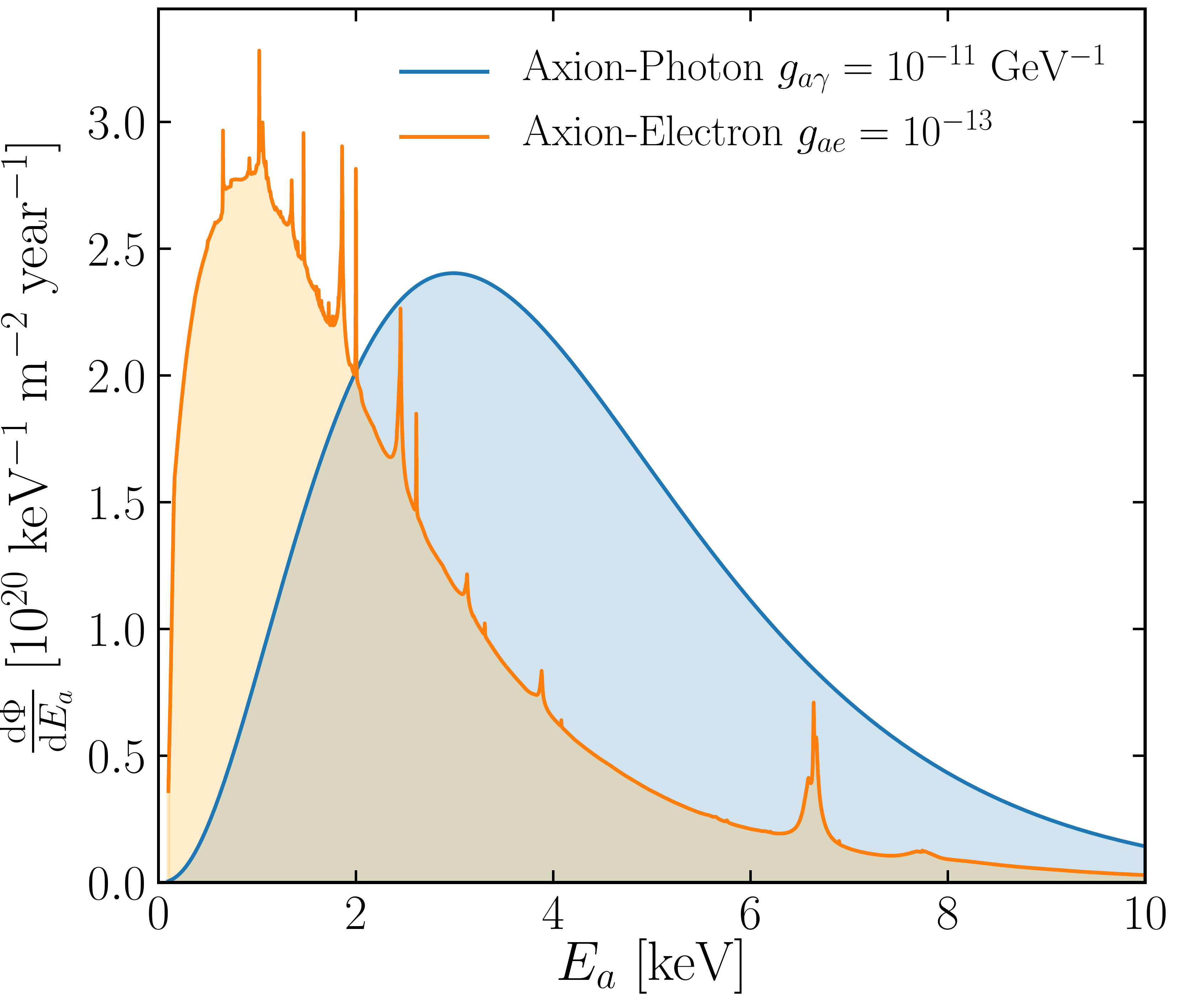}
\caption{\label{fig:fluxes} The solar axion flux expected on Earth and its components due to the axion-electron coupling (bremsstrahlung, Compton and axio-recombination) in orange and the axion-photon coupling (Primakoff) in blue. For this plot we assume $g_{a \gamma}=10^{-11}$ GeV$^{-1}$ and $g_{ae}=10^{-13}$.}
\end{figure}

\begin{figure*}
\centering
\includegraphics[width=0.99\textwidth]{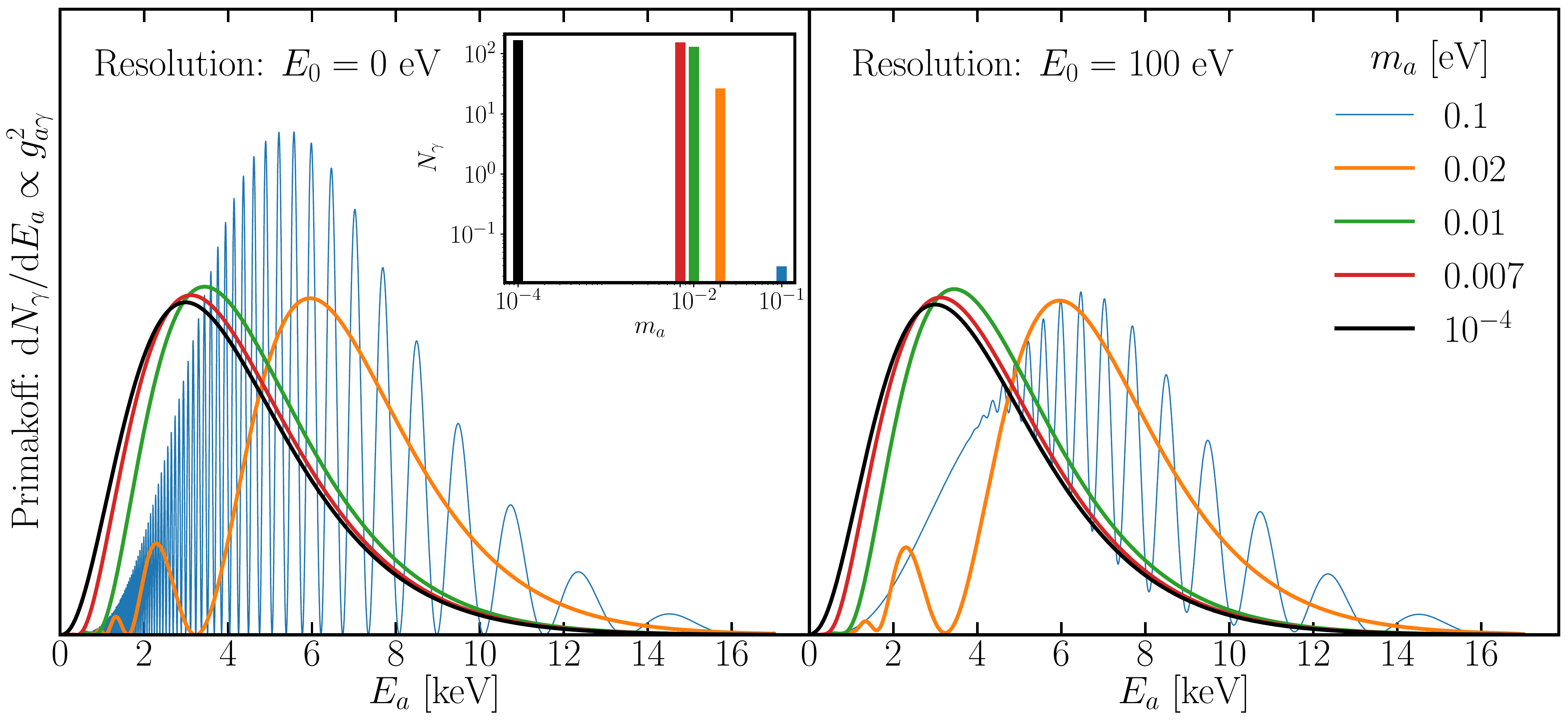}
\includegraphics[width=0.99\textwidth]{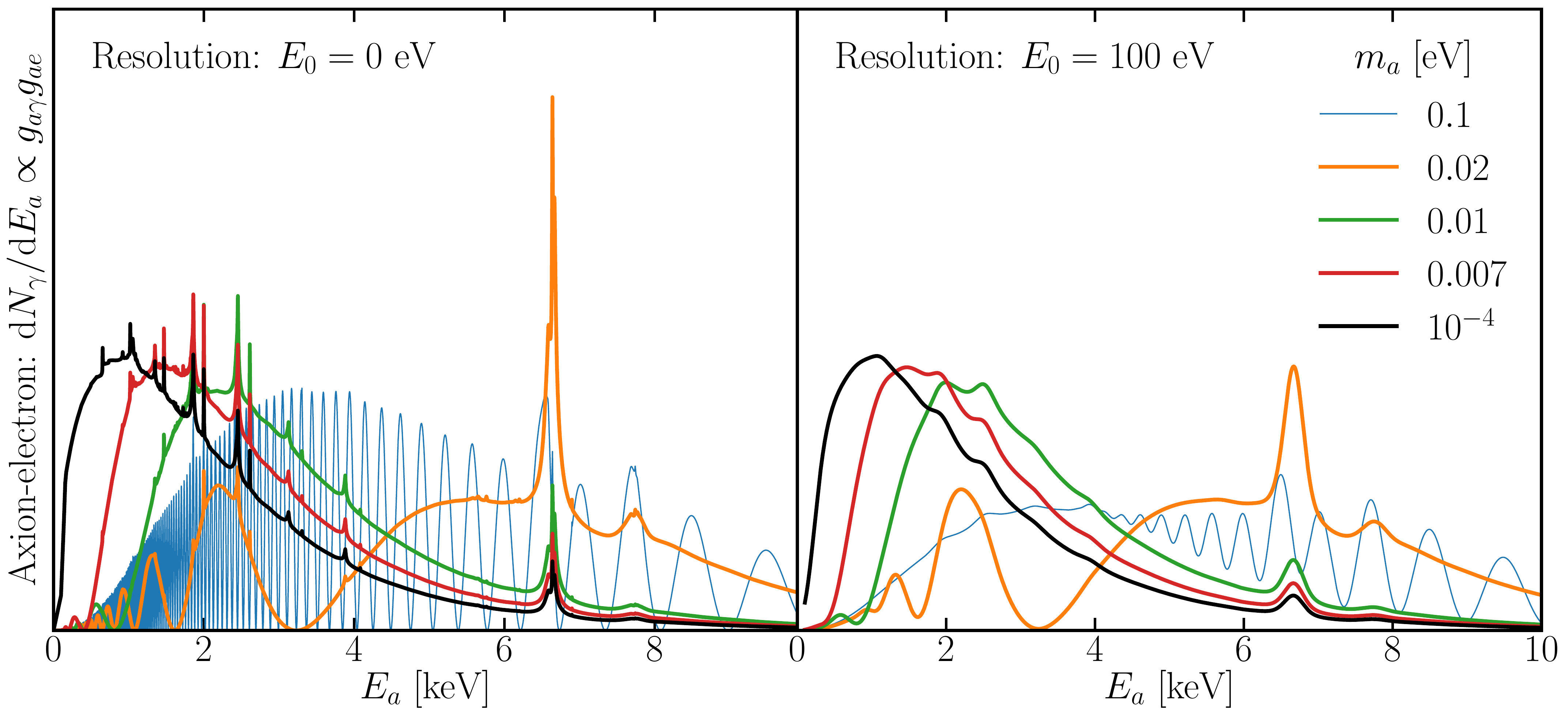}
\caption{Differential x-ray spectra as a function of energy due to solar axion conversion inside a 20 m long 2.5 T magnet. We display spectra for different values of the axion mass $m_a$ as well as for both the solar Primakoff ({\bf top}) and axion-electron ({\bf bottom}) fluxes. The left-hand panels in both cases show the underlying spectra, whereas the right-hand panels show the spectra after being convolved with a Gaussian energy resolution of width $E_0 = 100$ eV. For comparison, we have normalised all spectra to one. Instead we display in the {\bf inset}, the total integrated number of events $N_\gamma$ as a function of the five masses, assuming $g_{a\gamma} = 10^{-11}$ GeV$^{-1}$.}
\label{fig:spectra}
\end{figure*}

If the axion exists, then the Sun will be a factory, generating axions in its centre via several model-dependent processes. Given the most readily observable coupling---that to the photon $g_{a\gamma}$---the Sun will produce axions via Primakoff conversion. The differential flux at Earth due to this mechanism, $\Phi_{\rm P}$, can be parameterised as (assuming $E_a$ is always in units of keV):
\begin{equation}
\label{prima}
\frac{{\rm d} \Phi_{\rm P}}{{\rm d} E_{a}} =  \Phi_{\rm P10} \left( \frac{g_{a \gamma}}{10^{-10} \, \rm{ GeV}^{-1}} \right)^{2} \frac{E_{a}^{2.481}}{e^{E_{a}/1.205}} \, ,
\end{equation} 
where $\Phi_{\rm P10} = 6.02 \times 10^{10} \, {\rm cm}^{-2}\,{\rm s}^{-1}\,{\rm keV}^{-1}$. This flux is dominant in hadronic axion models like the KSVZ~\cite{Kim:1979if,Shifman:1979if}.

For nonhadronic models which possess a tree-level coupling to electrons, $g_{ae}$, the axion flux is instead dominated by three distinct processes: Compton scattering ($\Phi_{\rm C}$), bremsstrahlung ($\Phi_{\rm B}$) and atomic recombination and deexcitation ($\Phi_{\rm A}$)~\cite{Redondo:2013wwa}. The former two can be parameterised as,
\begin{equation}
\frac{{\rm d} \Phi_{\rm C} }{{\rm d} E_{a}}= \Phi_{\rm C13} \left( \frac{g_{ae}}{10^{-13}} \right)^{2} \frac{E_{a}^{2.987}}{e^{0.776E_{a}}} \, ,
\end{equation} 
where $\Phi_{\rm C13} = 13.314\times 10^6 \, {\rm cm}^{-2}\,{\rm s}^{-1}\,{\rm keV}^{-1}$, and
\begin{equation}
\frac{{\rm d} \Phi_{\rm B}}{{\rm d} E_{a}} = \Phi_{\rm B13}\left( \frac{g_{ae}}{10^{-13}} \right)^{2} \frac{E_{a}}{1+0.667E_{a}^{1.278}}e^{-0.77E_{a}} \, ,
\end{equation} 
where $\Phi_{\rm B13} = 26.311 \times 10^{8} \, {\rm cm}^{-2}\, {\rm s}^{-1}\,{\rm keV}^{-1}$. However $\Phi_{\rm A}$ cannot be efficiently or accurately parameterised so we have included it numerically using the simulation result of Ref.~\cite{Redondo:2013wwa}. Figure~\ref{fig:fluxes} compares the solar axion flux expected on Earth produced via the axion-photon (blue) and axion-electron (orange) coupling. In general the flux of the axion-electron processes is much more intense than the corresponding Primakoff flux, and is shifted to lower energies, peaking around 1~keV.

In concrete axion models the ratio of the axion-electron to Primakoff flux is prescribed. 
We have verified that in nonhadronic benchmark models like DFSZ I and II  the axion-electron flux tends to dominate throughout the whole spectrum,  while for the hadronic model KSVZ---where the electron coupling is generated radiatively---the Primakoff flux dominates. Therefore it is reasonable to study the axion-electron and Primakoff cases separately. Note however, that in some models like the Axi-majorons, both contributions can be of the same order. Details about these models and their couplings can be found in Ref.~\cite{Giannotti:2017hny}.

\subsection{Helioscopes}
A helioscope consists of a long magnetic bore pointed at the Sun with a collecting x-ray detector at one end. Independently of the mechanism of their production in the Sun, a helioscope relies on $g_{a\gamma}$ to convert the axion flux into photons. This leads to a signal proportional to $g_{a\gamma}^4$ for the Primakoff flux and $g_{ae}^2g_{a\gamma}^2$ for the axion-electron flux. The expected number of photons reaching a detector placed at the end of the bore is given by the integral
\begin{equation}\label{eq:Ngamma}
 N_{\gamma} = S\, t\, \int  dE_{a}\, \varepsilon_{\rm D}(E_a)\varepsilon_{\rm T}(E_a)\frac{{\rm d} \Phi_i}{{\rm d} E_{a}} \, P_{a \rightarrow \gamma}(E_a)  \, ,
\end{equation}
where $\frac{{\rm d} \Phi_i}{{\rm d} E_{a}}$ the axion flux due to process $i$ (i.e. P or  A+B+C), $P_{a \rightarrow \gamma}$ is the axion-photon conversion probability, $S$ the total cross-sectional area of the helioscope, and $t$ the measurement time. We parameterise two efficiency functions for the detector ($\varepsilon_{\rm D}$) and the telescope ($\varepsilon_{\rm T}$). 

The axion-photon conversion probability inside the magnet bore is (assuming a vacuum),
\begin{equation}
P_{a \rightarrow \gamma}(E_a) = \left(  \frac{g_{a \gamma}B}{q} \right)^{2} \sin^{2}\left( \frac{qL}{2} \right) \, ,
\label{Pvac}
\end{equation}
where $L$ is the magnet length, $B$ the magnetic field, and $q=m_{a}^{2}/(2E_{a})$ the axion-photon momentum transfer. The conversion probability is maximised when the axion and photon remain in phase over the length of the magnet, satisfying the coherence condition $qL < \pi$. As a result, the experimental sensitivity for a vacuum filled magnet bore is restricted to a range of axion masses, e.g. $m_a \lesssim 0.016$~eV for $L=20$~m and $\langle E_{a}\rangle=4.2$~keV.  

\begin{table}[t!]
\ra{1.3}
\begin{tabularx}{0.35\textwidth}{X|ll}
\hline\hline
Magnetic field	& $B$\quad & 2.5 T \\
Length	& $L$ \quad& 20 m \\
Total aperture area	& $S$ \quad& 2.26 m$^2$ \\
Measurement time	& $t$ \quad& 3 years\\
Telescope efficiency & $\varepsilon_{\rm T}$\quad & 0.8 \\
Detector efficiency & $\varepsilon_{\rm D}$ \quad& 0.7 \\
Energy resolution & $E_0$\quad & 10--200 eV\\
\hline\hline
\end{tabularx}
\caption{IAXO experimental configuration fixed in this study, as well as the range of energy resolutions we consider.}
\label{tab:IAXOparams}
\end{table}
For IAXO to achieve its stated sensitivity, the construction of a large-scale ($L \approx 20$~m) strong magnet with multiple bores is envisaged. The bores would have large aperture ($\sim 60$~cm diameter) and would be equipped with x-ray optics focusing the photons to a few spots of $\rm{mm}^{2}$ size. The signal areas would be imaged by ultralow background x-ray detectors like micromegas~\cite{Aznar:2015iia}, optimised for energies between $10$~keV down to well below $1$~keV~\cite{Garza:2016nty}. Additional detection technologies, like GridPix~\cite{Krieger:2018nit}, metallic magnetic calorimeters (MMCs)~\cite{Kempf:2013oca,Gastaldo:2012nv}, transition edge sensors  or silicon drift detectors are also under consideration, promising better energy threshold and resolution than the baseline Micromegas detectors. In particular, the most advanced MMCs have shown a 2 eV full width at half maximum energy resolution. So a suitable range for our energy resolutions to capture the importance of this experimental parameter while remaining realistic is $E_0  = $10--200 eV (where we quote $E_0$ as the standard deviation for the Gaussian smoothing kernel). The system would be capable of tracking the Sun for about $12$~h each day.

The procedure we follow in this work is based around extracting the mass dependence found in the axion signal Eq.\eqref{eq:Ngamma}. The numerical values we adopt for the experimental parameters entering this formula are summarised in Table~\ref{tab:IAXOparams}. We use the configuration anticipated in the IAXO conceptual design~\cite{Irastorza:2013dav, Armengaud:2014gea} which assumes eight bores (total $S=2.26$ m$^2$) and a 3 year total data-taking time. IAXO is considering several x-ray detector technologies for the focal planes of the telescope. So in order to keep the widest generality, we have convoluted Eq.\eqref{eq:Ngamma} with a Gaussian of width $E_0$ representing an energy resolution for the detectors. The energy threshold is assumed to be equal to $E_0$. Again following Refs.~\cite{Irastorza:2013dav, Armengaud:2014gea} we assume the telescope and detector efficiency functions are flat in energy, with $\varepsilon_{\rm T} = 0.8$ and $\varepsilon_{\rm D} = 0.7$ respectively. 

The background level in the IAXO detectors is expected to be extremely low \cite{Irastorza:2013dav, Armengaud:2014gea}, amounting to only a few counts in the signal region of interest over the full data-taking campaign. Since the detection of the axion mass will always require a larger number of signal events than this, we have assumed a zero background. It is also highly unlikely that the IAXO background will share any spectral properties with the signal, so including it would not have a significant impact on the results. 

\subsection{The axion mass signal}
One of the advantages of the helioscope technique is the capability to explore such a wide range of axion masses. For light axions which satisfy the coherence condition, the expected signal is viewed as independent of $m_{a}$. This means that a helioscope can set limits to arbitrarily small masses. Of course, this comes at the cost of having no ability to measure the mass. Even with a positive detection of an axion with a mass in this regime, only the coupling is strictly measurable. In fact, such an axion is not even distinguishable from a massless analogue particle. However the axion, per its definition, is not massless. If heavy enough, the coherence condition, which facilitates the axion detection for low masses, will be violated. The conventional wisdom for a helioscope is that for axions that are too massive the signal is destroyed by the oscillations brought about when $qL>\pi$ in the sin$^2$ term in the axion conversion probability~\eqref{Pvac}. But this is not immediately true. There is an intermediate regime, one in which the coherence condition is only mildly violated. This same oscillatory term will give rise to highly characteristic spectral oscillations that are strong enough to be measurable, but not so strong as to destroy the signal entirely. 

As an illustration, we show in Fig.~\ref{fig:spectra} the expected x-ray photon spectra ${\rm d}N_\gamma/{\rm d}E_a$, for values of $m_a$ between the effectively massless $10^{-4}$~eV (black line) and $10^{-1}$~eV (blue line). We show spectra observed from the axion-photon flux controlled by $g_{a\gamma}$ (upper row) and the axion-electron flux controlled by $g_{ae}$ (lower row). Notice also that in the $g_{a\gamma}$ case there are only very small differences in the spectra in the 2 orders of magnitude between $m_a = 10^{-4}$ and 0.01 eV (comparing the black and green lines). Clearly high statistics will be required for these lowest masses, meaning we can anticipate that the sensitivity of a helioscope to the axion mass will rise rapidly in coupling towards smaller $m_a$. This is true even in the limit of zero energy threshold and despite the fact that the number of events remains high down to low masses (see inset). In the right-hand panels of Fig.~\ref{fig:spectra} we also show the resulting spectra after a convolution with a Gaussian energy resolution function with an energy-independent width $E_0 = 100$ eV. A finite energy threshold or resolution we can foresee will play a greater importance at these lowest masses where the Gaussian will smooth out the very small low-energy oscillations that will be characteristic of a given value of $m_a$. In the case of heavier axions, for which the oscillations continue to higher energies, the finite resolution will likely be less important.

In principle one should also account for the systematic uncertainties on the theoretical spectrum, which come mostly from our imperfect understanding of plasma screening effects\footnote{An educated estimate outputs a maximum uncertainty of $\sim$30\% in the case of bremsstrahlung. This is certainly negligible for our purposes.}.  
In order to overcome this drawback, one can consider taking additional data with an artificially reduced magnet length, e.g. $L/2$. For the masses under consideration here, in which the coherence condition is only slightly violated ($qL\sim \pi$). Data taken from a half-magnet length would still enjoy the coherence condition $qL/2< \pi$, effectively amounting to a measurement of the reference zero-mass spectrum. This additional reference spectrum could be used to subtract off any underlying theory systematic. For the sake of simplicity, in the present study we have not followed such a procedure and are assuming our spectra are exempt from theoretical uncertainty. Nevertheless, in the event of a signal, a detailed study of its dependence on $L$ would be a mandatory cross-check to assess its axionic nature.

The distinction between the Primakoff and axion-electron fluxes (or a mixture thereof) will be another important task for IAXO should a detection be made. A similar spectral analysis to that which we describe here will be appropriate for this. However we do not focus on model discrimination here and deal mainly with the measurement of the mass given an assumption about the spectrum. This is in order to highlight several physical effects and requirements on the performance of the IAXO detectors. In addition, as mentioned previously, the reference massless spectrum is always obtainable from a shorter magnet length experiment. A similar recent work which approaches the problem from a model discrimination perspective can be found in Ref.~\cite{Jaeckel:2018mbn}.

\subsection{Higher axion masses}
Our study here deals only with axion masses for which the coherence condition is held ($m_a \lesssim 10^{-2}$ eV), or only mildly violated ($m_a \approx 10^{-2} - 10^{-1}$ eV). Since we do not consider higher masses than this in detail, before we proceed it is worth describing schematically how one would approach a similar analysis for them.

For axion masses above the coherence condition, axion-photon oscillations along the magnetic field destroy the signal. It is possible however to recover the condition by filling the bore with a buffer gas. The new medium provides an effective photon mass $m_\gamma = \sqrt{4\pi \alpha n_e/m_e}$, where $n_e$ and $m_e$ are the electron density and mass and $\alpha$ is the fine structure constant. When the axion mass matches this value, the axion-photon momentum transfer vanishes and axions can convert coherently across the magnet length.

In this setup, the axion-photon conversion probability takes the form~\cite{vanBibber:1988ge},
\begin{align}
P_{a \rightarrow \gamma} =& \left(  \frac{g_{a \gamma}B}{2} \right)^{2}  \frac{1}{q^{2}+ \Gamma^{2}/4} \nonumber \\
&\times  \left( 1 + e^{-\Gamma L} - 2 \, e^{-\Gamma L/2} \cos(qL) \right) \, ,
\end{align} 
where $q=|m_{a}^{2}-m_{\gamma}^{2}|/(2E_{a})$ and $\Gamma$ is the inverse absorption length for photons in a buffer gas. A search over a range of masses can then be performed by tuning the buffer gas density. 

In the gas scanning mode the coherence is restored in a range of masses $\Delta m_a \simeq \pi/2 \sqrt{4 E/L}$~\cite{Arik:2008mq} ($\sim 0.02$ eV for IAXO using $E_a\sim 4$ keV), which is set as the natural $\Delta m_a$ between consecutive pressure settings. This gives the precision with which the axion mass can be measured from the $\mathcal{O}(1)$ difference in the total x-ray flux at neighbouring steps. For heavier axions, the measurement of $m_a$ in this mode is then already essentially complete when the axion itself is detected. Naturally, if one wanted higher precision this would be achieved by measuring smaller flux differences between steps, requiring higher statistics. Experimentally this would involve monitoring the pressure to better precision than CAST and taking multiple pressure settings. This could allow one to measure the axion mass even if it is smaller than the natural $\Delta m_a \sim 0.02$ eV, but only if the axion coupling is large enough. Nevertheless, since the signal differences between consecutive pressure steps will be small, a spectral analysis would be needed. 
In the rest of the paper we show how a much simpler determination of the axion mass can be obtained already in the vacuum phase by measuring the spectral distortions of the conversion probability. This involves a similar level of sophistication in the analysis but much less in the experiment itself, and has the advantage that it can be done already with the vacuum mode data. 

\section{Statistical methodology}\label{sec:stats}
\begin{figure}
\includegraphics[width=0.49\textwidth] {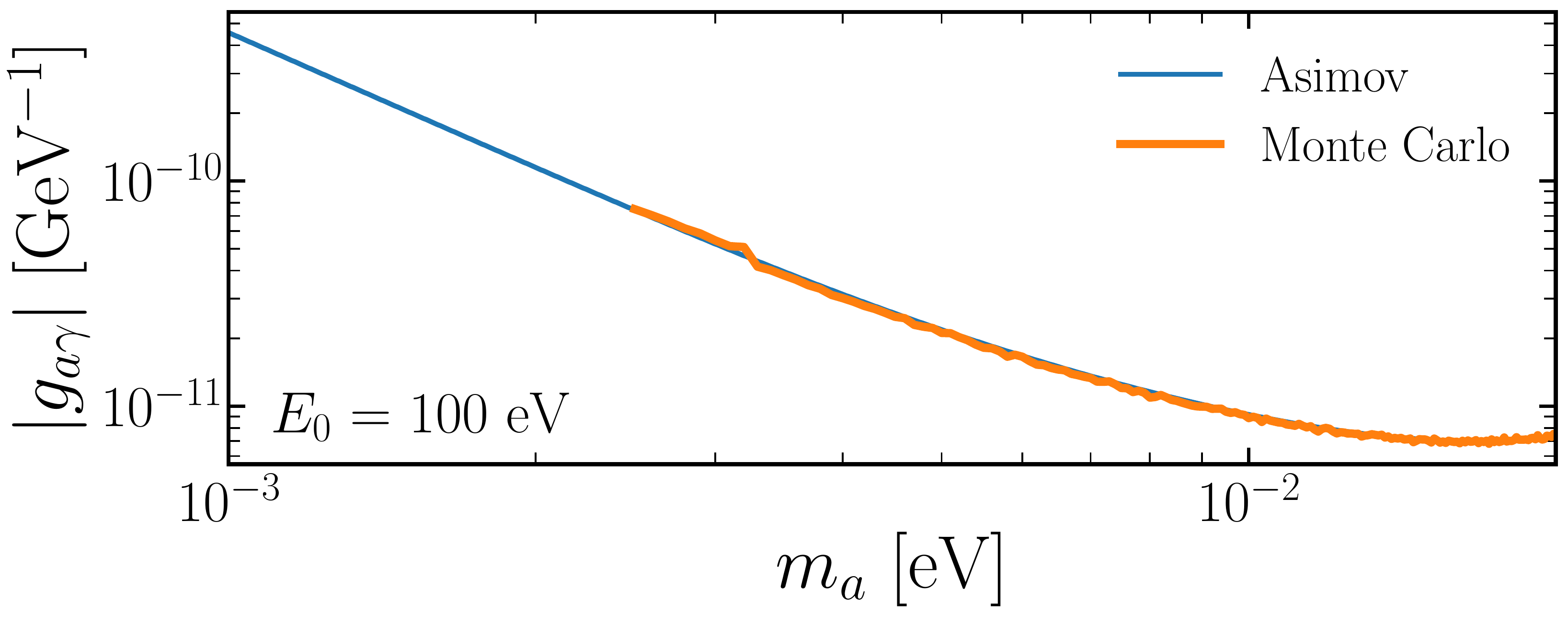}
\caption{Comparison between one of the median mass discovery limits obtained via the Asimov data formalism (blue) and the same when computed via a full Monte Carlo simulation of the test statistic distribution (orange).}\label{fig:asimov}
\end{figure}

\begin{figure*}
\includegraphics[width=0.49\textwidth] {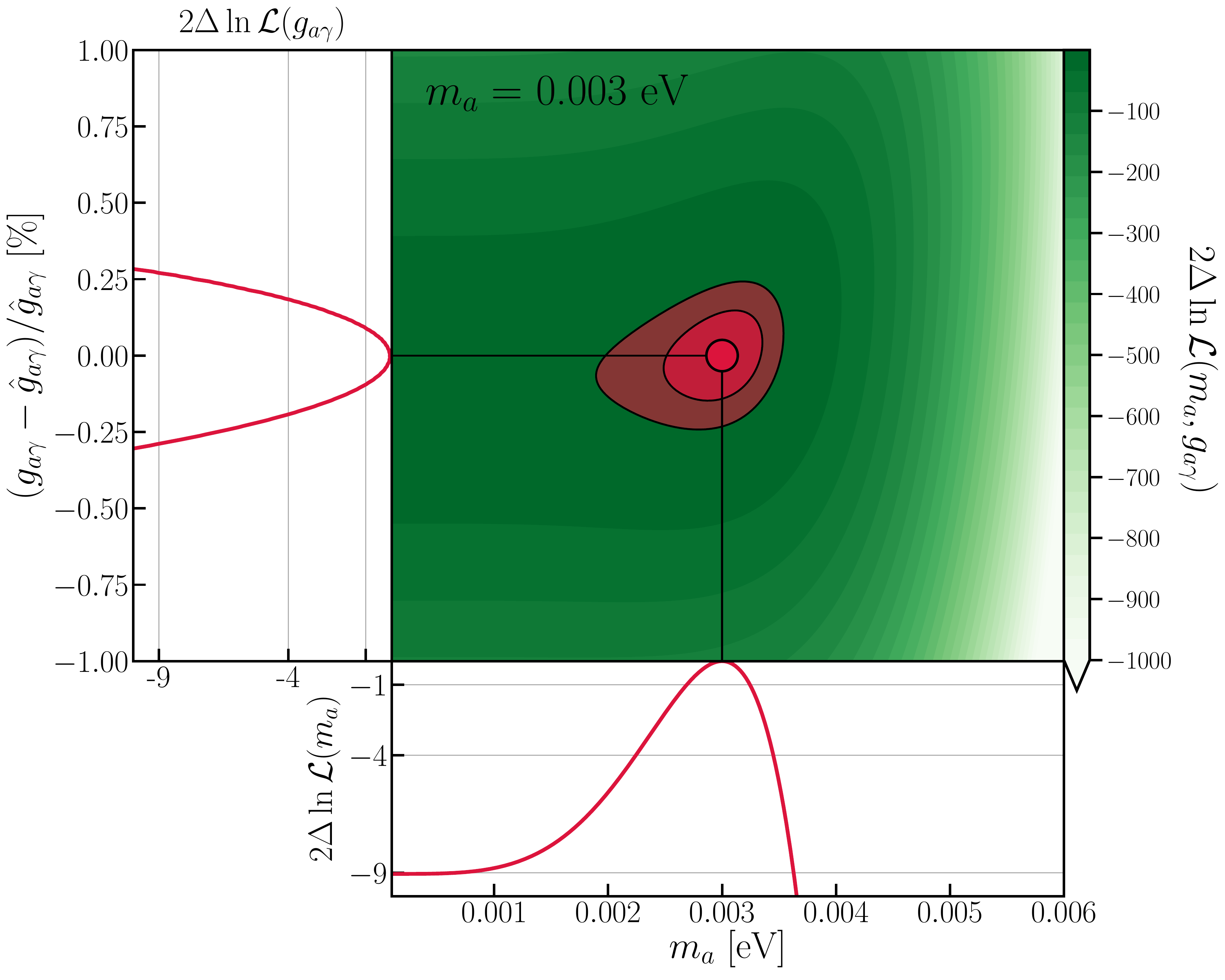}
\includegraphics[width=0.49\textwidth] {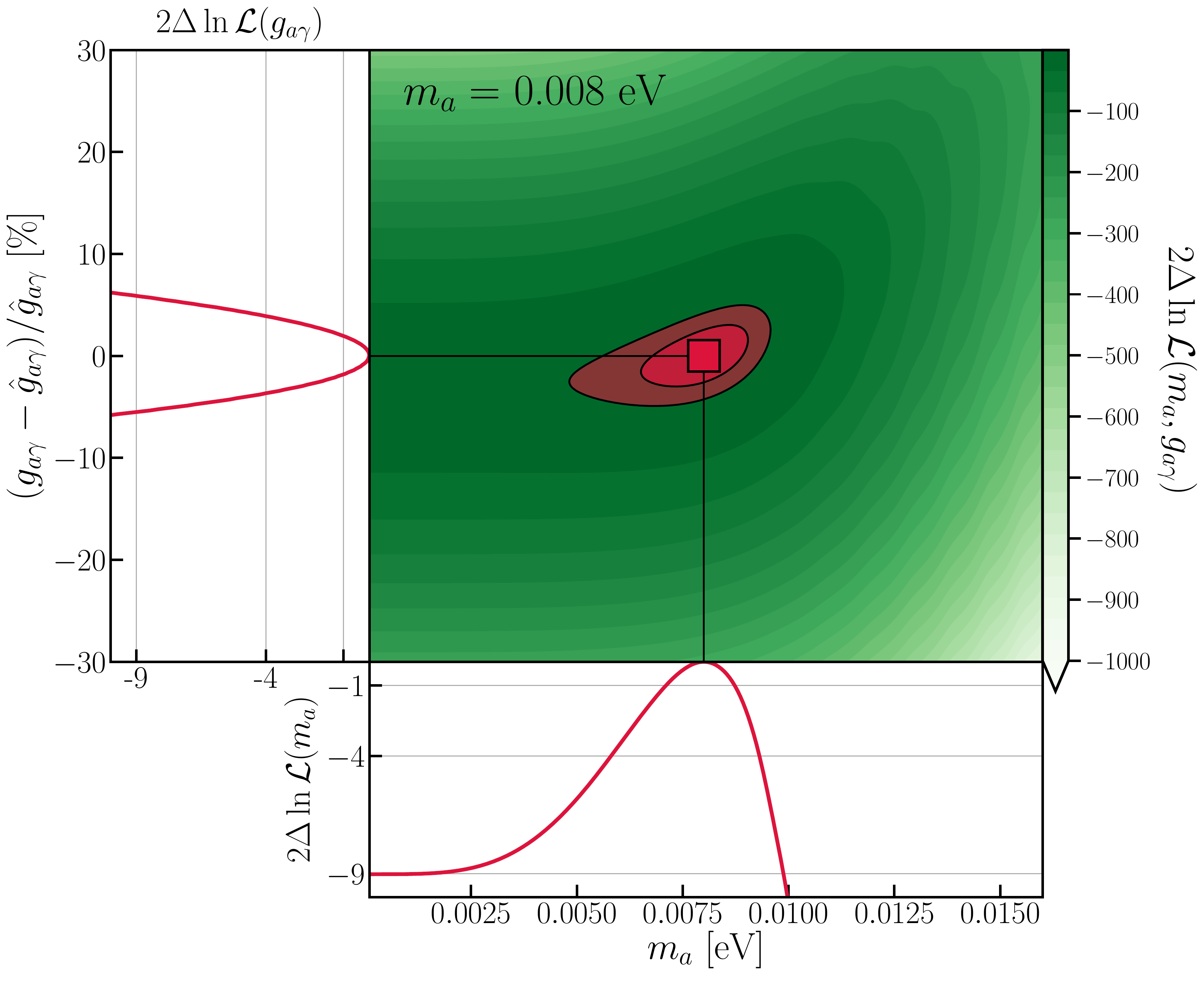}
\includegraphics[width=0.49\textwidth] {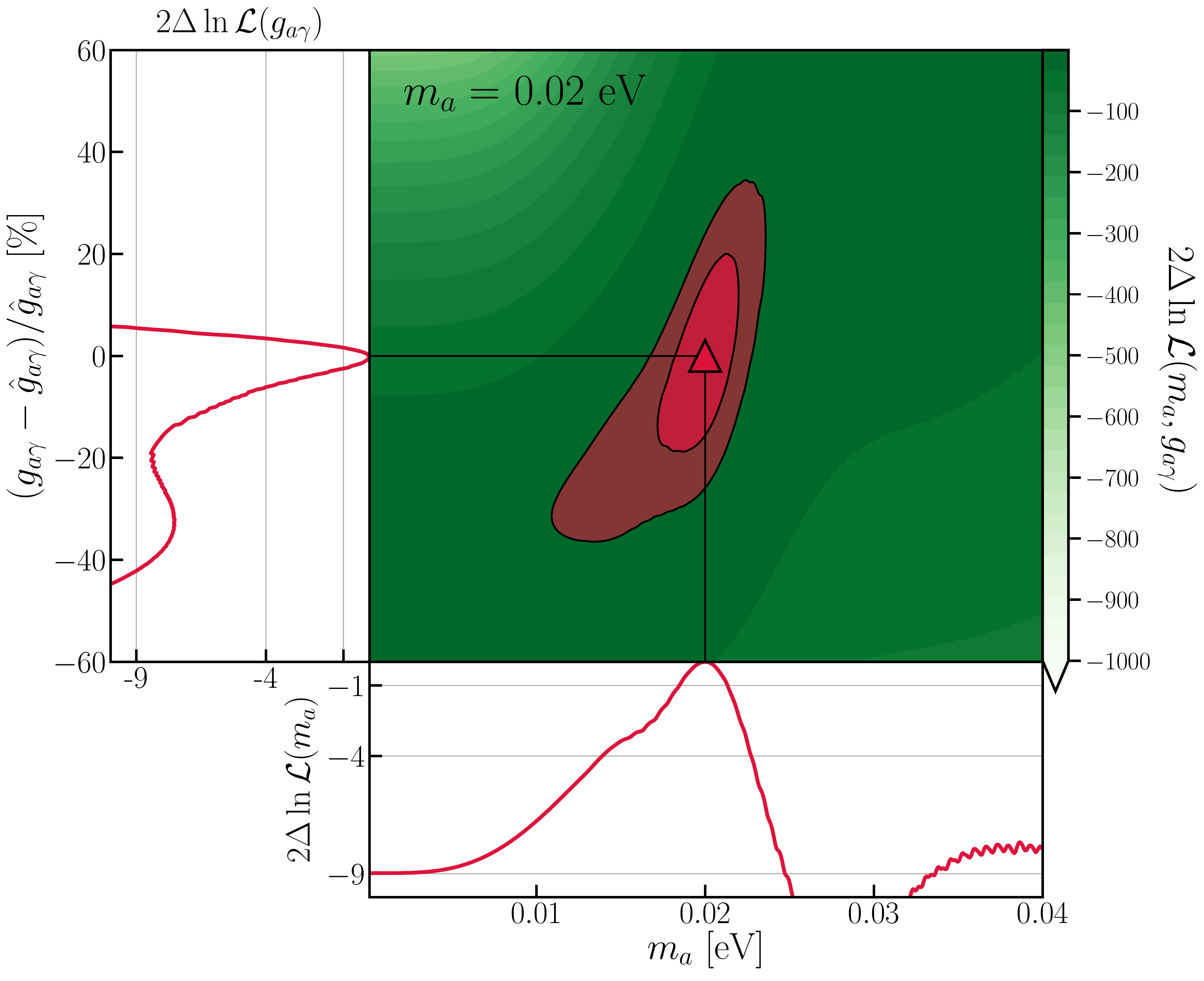}
\includegraphics[width=0.49\textwidth] {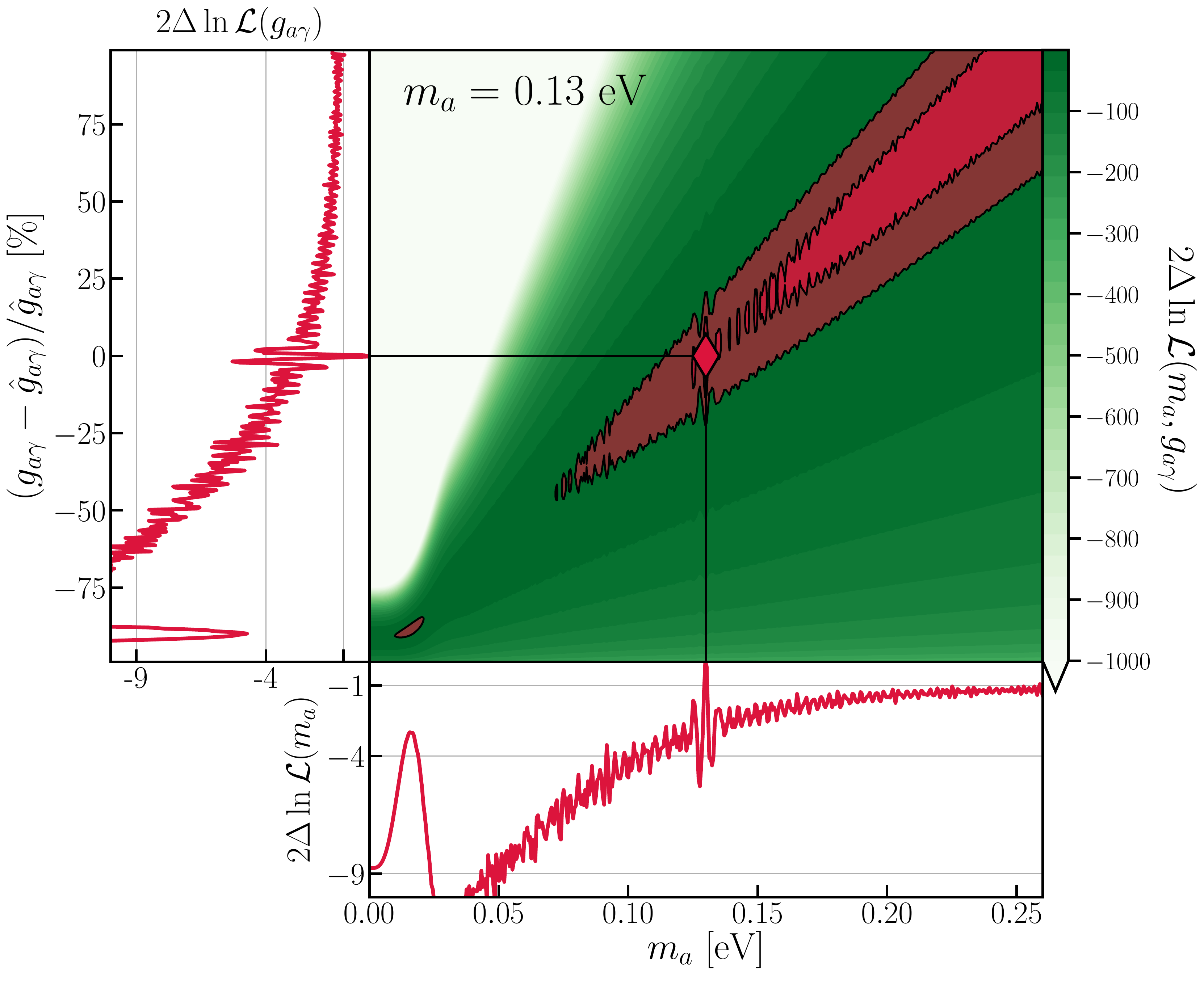}
\caption{\label{fig:Like} Two- and one-dimensional profile log-likelihood ratios for four input axion masses, where $g_{a\gamma}$ has been chosen to yield a 3$\sigma$ discrimination of the mass from $m_a = 0$, assuming a Primakoff dominated axion flux. In each case we plot the difference between the likelihood value and the maximum likelihood, which, since we are using Asimov data, is always correctly located at the input parameter values (indicated by straight lines and a red marker). In the two-dimensional likelihood we also show the 1$\sigma$ and 2$\sigma$ enclosed contours on both parameters.}
\end{figure*}

Our statistical methodology is based on the popular profile likelihood ratio test, in common use for deriving discovery and exclusion limits in particle physics experiments. Our hypothesis test compares the massive axion model
$\mathcal{M}_{m_a \neq 0}$ with two parameters $(m_a, g)$ against the massless model, $\mathcal{M}_{m_a=0}$ with only one parameter $(g)$. Here we use the generic $g$ to imply either $g_{a\gamma}$ or $\sqrt{g_{a\gamma} g_{ae}}$. 

We can construct the profile likelihood ratio between the two models,
\begin{equation}\label{eq:likelihood-ratio}
\Lambda = \frac{ \mathcal{L} (0,\hat{\hat{g}} ) }{\mathcal{L} (\hat{m}_a,\hat{g})  }\, ,
\end{equation}
where~$\mathcal{L}$ is a likelihood function which is maximised at
$\hat{\hat{g}}$ when $m_a$ is set to zero, and $(\hat{m}_a,\hat{g})$ when $m_a$ is free. We next define the profile likelihood ratio test statistic,
		\begin{equation}\label{eq:q0}
			q_0 = \left\{ \begin{array}{rl}
			-2\ln \Lambda  & \, \, \hat{m}_a>0 \,,\\
			0  & \, \, \hat{m}_a<0 \,.
			\end{array} \right. 
		\end{equation}
Since the two models differ by the fixing of one parameter and the hypothesis is being compared against a parameter at the boundary of the allowed space, Chernoff's theorem~\cite{Chernoff:1954eli} holds. This is a generalisation of Wilk's theorem and states that the statistic $q_0$ is asymptotically distributed according to $\frac{1}{2}\chi^2_{1}+\frac{1}{2}\delta(0)$ when the $\mathcal{M}_{m_a=0}$ hypothesis is true. The consequence of this that is useful for us is that the significance of the signal from a massive axion when tested against the massless hypothesis is simply $\sqrt{q_0}$. See Ref.~\cite{Cowan:2010js} for a detailed discussion of the use of these asymptotic formulae.

We use a binned likelihood for $\mathcal{L}$ so that we can employ the Asimov asymptotic limit for the test~\cite{Cowan:2010js} in which the number of events in each bin is set equal to the expectation. The value of $q_0$ yielded when applying the profile likelihood ratio test on this data approaches the median value that would be obtained from many Monte Carlo realisations. This method can therefore straightforwardly provide us with the median discovery projections, greatly saving on computational cost. Nevertheless, we have verified the Asimov formalism with Monte Carlo simulations and find very good agreement as can be seen in Fig.~\ref{fig:asimov}.

The Asimov dataset technique is in common use in studies like this one for facilitating the fast computation of asymptotic limits on particle physics properties, e.g Refs.~\cite{OHare:2018trr,Knirck:2018knd,Edwards:2018lsl,Foster:2017hbq}. It is not surprising that we obtain good agreement since an Asimov test statistic converges on the median Monte Carlo result rapidly with few events for strongly Gaussian or Poissonian likelihoods. In any case we only want to estimate the median value of the test statistic distribution and hence compute the ``typical'' limit IAXO might set. For the median value the relationship between the Asimov and Monte Carlo results is asymptotically exact. One caveat is that we must use typically in excess of 200 bins between $[E_0,20]$ keV. This is required for our Asimov limit to match the Monte Carlo results using an event-by-event unbinned likelihood. Since very fine energy information is needed to make measurements of the mass we anticipate that the most powerful approach in a real experimental scenario would be to resort back to an unbinned likelihood.

The binned likelihood that enters Eq.~\eqref{eq:likelihood-ratio} is the product of the Poisson probability distribution function $\mathscr{P}$ for $N_{\rm obs}$ x-rays, given an expected number $N_{\rm exp}$, 
\begin{equation}
\label{eq:Like}
\mathcal{L}(m_a,g) = \prod_{i=1}^{ N_{\rm bins}}\mathscr{P}\left[N_\textrm{obs}^i \bigg| N^i_{\rm exp}(m_a,\,g) \right] \, .
\end{equation}
Since we are background free the expected number of events follows Eq.\eqref{eq:Ngamma}. To aid our discussion later we will write this schematically as
\begin{align}
\label{mui}
N_{\rm exp}^i(m_a,g) =& \left(\frac{g}{10^{-10}}\right)^4 \nonumber \\
&\times \int_{E^i_a}^{E^{i+1}_a} \textrm{d}E_a \frac{\textrm{d} \Fa}{\textrm{d}E_a} p(m_a) \nonumber \\
& \equiv g^4 \mathcal{N}^i(m_a) \, .
\end{align}
The function $\textrm{d}\Fa/\textrm{d}E_a$ is the x-ray spectral fluence with the coupling constant ($\gag$ or $\sqrt{\gag g_{ae}}$) fixed at $10^{-10}$ (GeV$^{-1}$ or GeV$^{-1/2}$). The utility of this object is that it is independent of any axionic parameters for both the Primakoff and axion-electron fluxes. Instead all of the axion dependence is stored in a coefficient and a form factor for the conversion probability\footnote{We use the definition $\sinc\,x = \frac{\sin{x}}{x}$.}, $p(m_a)=\sinc^2(m_a^2L/4E_a)$.  

Before we use this test to calculate the discovery limit on the axion mass, we can gain some intuition by illustrating the shape of the likelihood introduced above for different axion masses. In Fig.~\ref{fig:Like} we show the shape of the likelihood for four values of $m_a$ across our range of interest. In each case the coupling is chosen so that the likelihood ratio test statistic for the correct value of the axion mass is equal to 9 (which corresponds to a 3$\sigma$ detection of a nonzero mass, as is derived in the next section). The left-hand side and bottom panels in each of the four cases shows the one-dimensional profile likelihood for $g_{a\gamma}$ and $m_a$ respectively. One can notice that for the lower masses the likelihood is smooth and the axion parameters would be very well reconstructed. However for the largest mass shown here, $m_a = 0.1$ eV, the x-ray spectrum is highly oscillatory and the shape of the likelihood reflects this. Since we have used Asimov data, the maximum likelihood is always correctly located at the true values. However in a real dataset which will suffer Poissonian fluctuations, strong biases may be possible, leading to spurious reconstructions of the axion mass with confidence intervals not enclosing the true value.

\section{Results}\label{sec:results}
\subsection{Mass discovery}
\begin{figure*}
\includegraphics[width=0.49\textwidth] {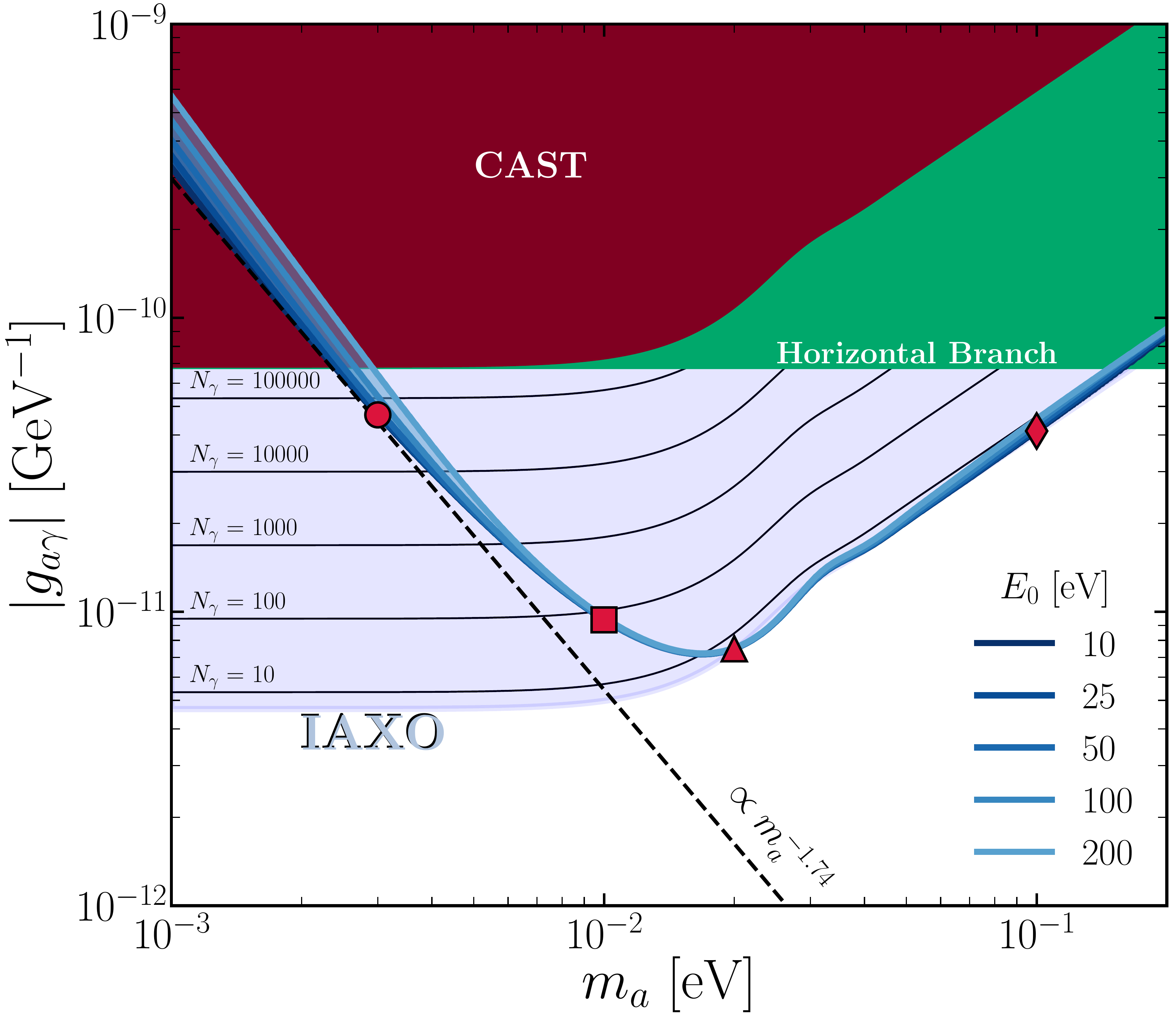}
\includegraphics[width=0.49\textwidth] {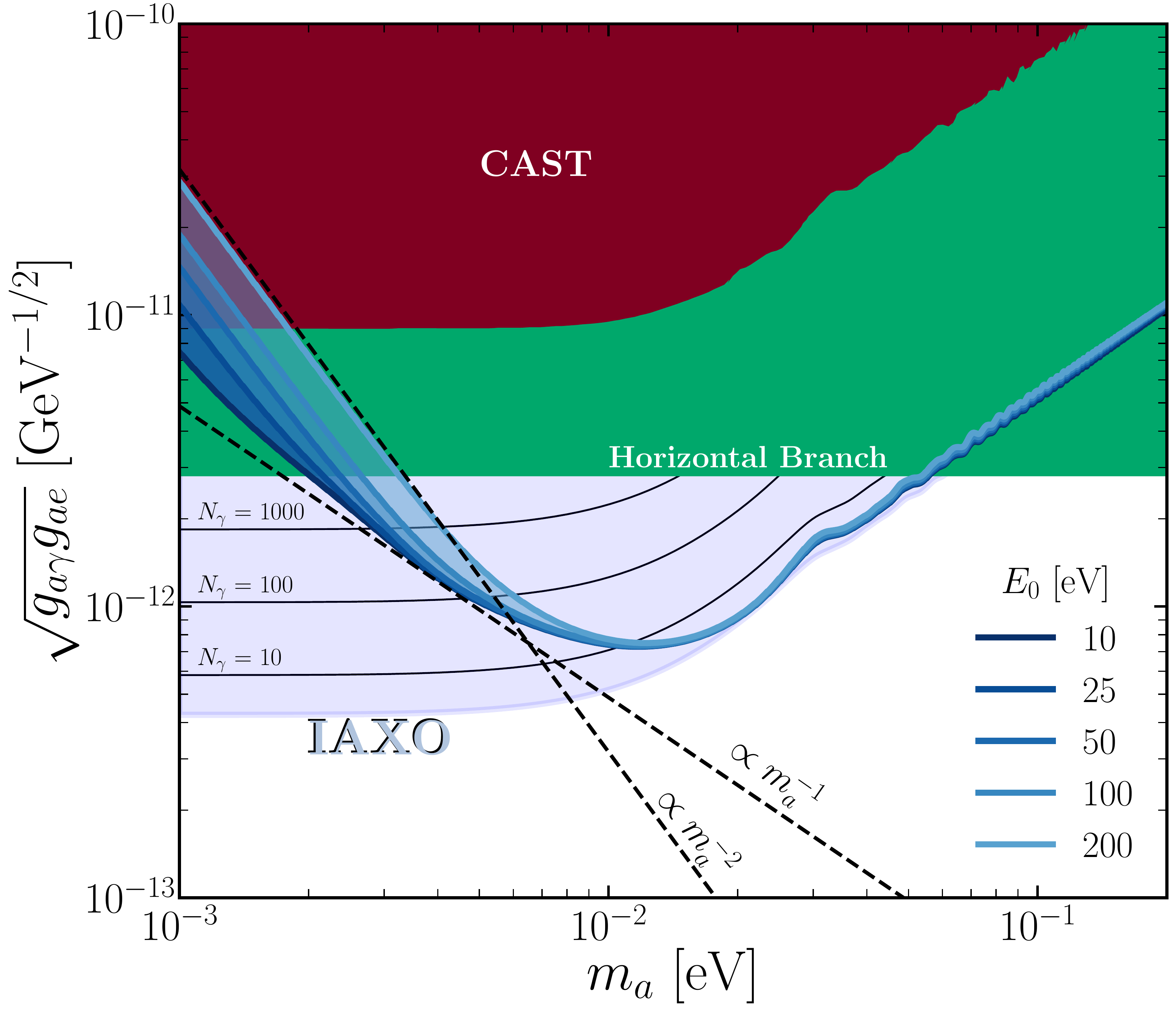}
\caption{\label{fig:result} Median discovery limits for determining a massive axion to $3\sigma$ in terms of the coupling to photons ({\bf left}) and electrons ({\bf right}). In each we plot from dark to light blue the discovery limits for increasing energy resolution $E_0$. The lightest blue region shows the sensitivity of IAXO to exclude $g_{a\gamma}$ or $\sqrt{g_{a\gamma}g_{ae}}$. The black lines indicate contours in these spaces which give constant numbers of x-rays $N_\gamma$. The symbols on the left-hand plot indicate the example values of the mass and coupling used for the four panels of Fig.~\ref{fig:Like}. The red regions indicate the exclusion limits from CAST on $g_{a\gamma}$~\cite{Anastassopoulos:2017ftl}, and $\sqrt{g_{ae}g_{a\gamma}}$~\cite{Barth:2013sma}, and the green indicates the limits on the same, from horizontal branch stars: Refs.~\cite{Ayala:2014pea} and \cite{Giannotti:2015kwo} respectively.}
\end{figure*}
We wish to calculate the minimum value of $g_{a\gamma}$ or $g_{a\gamma}g_{ae}$ as a function of $m_a$ for which IAXO can determine that the axion is not massless. We can display this by defining a ``mass discovery limit'' $g_{\rm disc}(m_a)$ to be the median coupling as a function of $m_a$ for which the mass can be distinguished from zero at $3\sigma$. Practically this requires us to calculate the smallest coupling for which $q_0 = 9$ when the test is applied to a set of Asimov data. A major advantage of the background and systematic-free likelihood is that this process can be done semianalytically and will lend us some insight into the impact of the threshold and energy resolution.

Since the coupling constant only enters as a multiplicative factor, the maximum likelihood estimators for the parameter $g$ when the likelihood is fixed at a particular value of $m_a$ can be calculated as simply,
\begin{align}
g_{\rm prof}^4(m_a) &= \frac{\sum_i N_{\rm obs}^i}{\sum_i \mathcal{N}^i(m_a)} \nonumber \\ &\equiv \, \mathcal{D}(m_a)\, .
\end{align}
When we set the mass to zero in this formula we get the maximum likelihood estimator for the massless likelihood, i.e. $g_{\rm prof}(0) = \hat{\hat{g}}$. Under the Asimov approximation we are guaranteed that $\hat{g}$ and $\hat{m}_a$ are the true values of the coupling and mass, so we write $N_{\rm obs} = \hat{g} \mathcal{N}(\hat{m}_a)$. Evaluating Eq.~\eqref{eq:q0} when set to nine for a $3\sigma$ result we get,
\begin{equation}
\label{sensitivitynonzeromass}
g_{\rm disc}(m_a) = 10^{-10}\,\(\frac{9}{2\sum_i \mathcal{N}^i(m_a)\log\frac{\mathcal{N}^i(m_a)}{\mathcal{D}(m_a)\mathcal{N}^i(0)}}\)^{\frac{1}{4}} \, ,
\end{equation}
where the dimensions of $g_{\rm disc}$ follow the dimensions of either $g_{a\gamma}$ or $\sqrt{g_{a\gamma}g_{ae}}$.

The lines corresponding to this condition are shown in Fig.~\ref{fig:result}, for both $g_{a\gamma}$ and $\sqrt{g_{a\gamma} g_{ae}}$. We show sets of curves corresponding to a range of energy thresholds/resolutions, $E_0$. The effect of the energy resolution is more pronounced in the latter case, as the spectrum is centred at lower energies. Generally, as one would expect, a higher energy threshold requires a stronger coupling in order to reach the same sensitivity. However only for the highest value considered here (200 eV) is the difference extended up to masses of $0.01$~eV, while the remaining cases converge below $m_a \sim~0.006$~eV. The principal cause for the differences between these curves present at low masses turns out to be the finite threshold at $E_0$, rather than the loss in spectral information by the energy resolution, as we now discuss.

The scaling towards small masses can be understood from considering the distribution of probabilities that make up the likelihood function. Comparing our discovery limits with the contours of constant $N_\gamma$ (black lines in Fig.~\ref{fig:result}), clearly when $m_a$ is small we are in a very high statistics regime. It is suitable then to approximate the likelihood with a product of Gaussian probabilities with standard deviation $\sqrt{N_{\rm exp}}$,
\begin{equation}
\mathcal{L}(m_a,g) \simeq \prod_{i=1}^{N_{\rm bins}} \frac{\left(N^i_{\rm obs} - N^i_{\rm exp}(m_a,g)\right)^2}{2 N^i_{\rm exp}(m_a,g)} \, .
\end{equation}
We take the further approximation of very small bins to convert these sums into integrals. Under Asimov data these approximations result in the following formula for the test statistic,
\begin{equation}
q_0 = \left(\frac{g}{10^{-10}}\right)^4 \int \textrm{d}E_a \frac{\textrm{d} \Fa}{\textrm{d}E_a} \frac{\(\mathcal{D}(\m)-p(\m)\)^2}{\mathcal{D}(\m)} \, ,
\end{equation}
where we have used the fact that $N^i_{\rm obs} = N^i_{\rm exp}(\hat{g},\hat{m}_a) = \hat{g} p(\hat{m}_a) \mathcal{N}^i(0)$.

Next we can use the fact that $\mathcal{D}(\m)\simeq 1$ for low values of $m_a$ because the total sum over bins is relatively mass insensitive, i.e. $\sum N_i(\m) \sim \sum N_i(0)$. This simplifies the expression further to,
\begin{equation}
\label{signi}
q_0 = \left(\frac{g}{10^{-10}}\right)^4 \int \textrm{d}E_a \frac{\textrm{d} \Fa}{\textrm{d} E_a} \[1-\sinc^2\(\frac{\m^2 L}{4 E_a}\)\]^2. 
\end{equation}
With the test statistic expressed in this way we can see that the power to determine the axion mass in fact scales with $[1-p(\m)]^2$. This is in contrast to the total number of events, (i.e. the power to determine the axion coupling), which scales with $p(m_a)$.  This is in accordance with the result of Fig.~\ref{fig:result} where the coupling exclusion limit of IAXO plateaus towards small masses since $p(m_a)\rightarrow 1$, but the mass discovery limit sharply increases since $[1-p(\m)]^2 \rightarrow 0$.

To understand more precisely this scaling in the mass discovery limit we can look at the behaviour of this function of $m_a$ with energy,  
\begin{equation}\label{eq:sincscaling}
\[1-\sinc^2\(\frac{\m^2 L}{4 E_a}\)\]^2 \sim \bb{1 & E_a< \frac{4}{9^\frac{1}{4} \m^2 L}}{\frac{1}{9}\(\frac{\m^2 L}{4 E_a}\)^4 & E_a> \frac{4}{9^\frac{1}{4} \m^2 L}} \, .
\end{equation}
Taking the Primakoff flux example first, we use the fact that its energy dependence is
\begin{equation}
\frac{\textrm{d}N_\gamma}{\textrm{d}E_a} \propto \frac{\textrm{d}\Phi_{\rm P}}{\textrm{d}E_a} \propto \frac{ E_a^{\beta} }{e^{E_a/1.205}}, \quad \text{with}\quad \beta=2.481 \, .
\end{equation}
Combining the two scaling regimes of Eq.\eqref{eq:sincscaling} tells us that the integrand of Eq.~\eqref{signi} must grow initially as $E_a^{\beta+1}$, but reaches a peak around 
\begin{equation}
E_{\rm \chi}\simeq  \frac{\m^2 L}{4} \, ,
\end{equation}
only then to decrease as $E_a^{1+\beta-4}$. Above the peak each log-interval contributes to the integral following a weak power law, $E_a^{\beta-3}\sim E_a^{0.51}$. So $q_0$ must be primarily influenced by energies around and above $E_{\chi}$. The signal at energies below the peak is rendered relatively unimportant for determining the mass. Approximating Eq.\eqref{signi} as just the integrand above the peak, we find 
\begin{align}
q_0 &\simeq  \left(\frac{g_{a\gamma}}{10^{-10}}\right)^4 \Phi_{\rm P10} \int^\infty_{E_{\chi}} dE_e E_a^{\beta} \frac{1}{9}\(\frac{\m^2L}{4 E_a}\)^4 \nonumber \\
&= \left(\frac{g_{a\gamma}}{10^{-10}}\right)^4\Phi_{\rm P10} \frac{(\m^2 L)^4}{2304(3-\beta)}\frac{1}{E_{\rm max}^{3-\beta}} \nonumber \\
&=  \left(\frac{g_{a\gamma}}{10^{-10}}\right)^4 \Phi_{\rm P10} \frac{4^{3-\beta}}{2304(3-\beta)}(\m^2 L)^{\beta+1} \, .
\end{align}
So we can see the following trend must follow, 
\be
g^{\rm disc}_{a\gamma} \propto q_0^{1/4} \propto \m^\frac{{\beta+1}}{2}\sim \m^{-1.74} \, ,
\ee
which reproduces the scaling at low masses of Fig.~\ref{fig:result} (left).

We can now understand the effect of the energy threshold which is cutting the integral at $E_0$. As long as $E_0<E_\chi$, the test statistic and hence the discovery limit will be unaffected. However, if $E_0>E_\chi$, the threshold removes a large contribution to $q_0$ and the sensitivity will suffer. Indeed we find
\begin{equation}
q_0 = g_{a\gamma}^4 \Phi_{\rm P} \frac{(\m^2 L)^4}{2304(3-\beta)}\frac{1}{E_0^{3-\beta}} \, .
\end{equation}
This quantity is only a factor of $(E_\chi/E_0)^{3-\beta}$ smaller than the zero-threshold case, so the mass discovery limit can only relax by $(E_0/E_\chi)^\frac{3-\beta}{4}$. 
Note that in this regime, the scaling of the discovery limit will change slightly to, 
\begin{equation}
g_{a\gamma}^{\rm disc} \propto \frac{1}{\m^2}, 
\end{equation}
which is in agreement with our results of Fig.~\ref{fig:result}, where one can see that higher thresholds have steeper inclines.

The axion-electron result follows a similar procedure. The only change is the shape of the axion flux, which at low energies goes as,   
\be
\frac{\textrm{d}\Phi_{\rm B}}{\textrm{d}E_a} \propto \frac{ E_a e^{-0.77 E_a}}{1+0.667 E_a^{1.278}} \sim E_a ,
\ee
so $\beta=1$. Indeed, our results approach this scaling as $E_0\rightarrow 0$. The integrand above the peak in this case decreases faster, $\propto E_a^{-3}$, leading to a greater sensitivity to the value of $E_0$. The curves for increasing $E_0$ in Fig.~\ref{fig:result} (right) approach the $m_a^{-2}$ scaling much more rapidly than the $g_{a\gamma}$ limits.

\subsection{Mass estimation}
\begin{figure}
\includegraphics[width=0.49\textwidth] {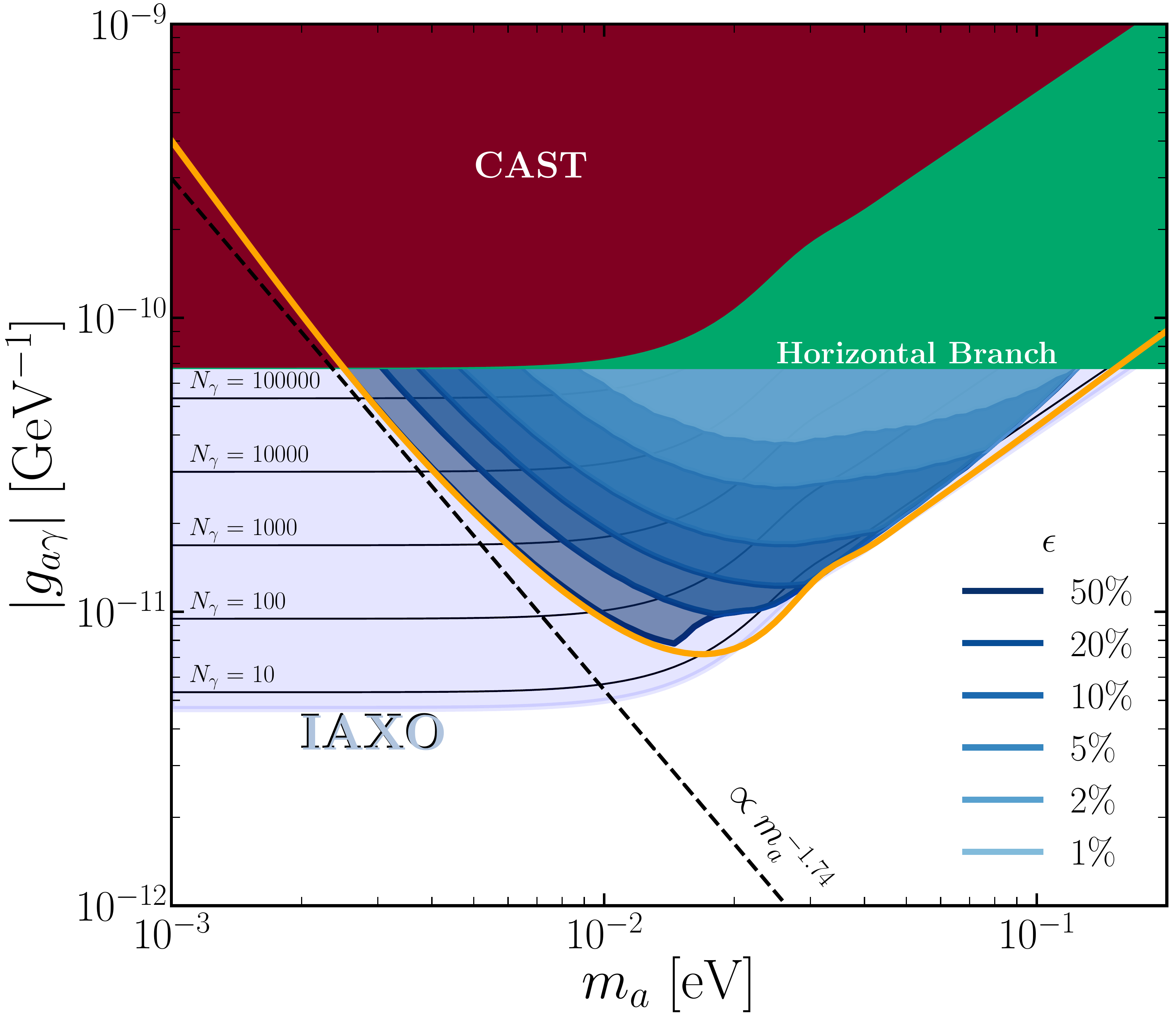}
\caption{\label{fig:massestimation} Limiting values of the axion-photon coupling that permit IAXO to obtain a 2$\sigma$ confidence interval around the axion mass, which is at most $\epsilon$ away from the true value. We show results for $\epsilon = $ 1\% to 50\%. We assume an energy resolution of 50 eV here.  For reference we include the mass discovery limit originally shown in Fig.~\ref{fig:result} (left) as an orange line.}
\end{figure}
As well as knowing how well the mass can be distinguished from zero, we also wish to estimate how well the mass itself can be measured. To do this we consider the one-dimensional profile likelihood function,
\begin{equation}
\mathcal{L}_1(m_a) = \mathcal{L}\left(g_{\rm prof}(m_a),\, m_a \right) \, ,
\end{equation} 
which is found by taking the likelihood defined previously, fixing a value of $m_a$, and finding the maximum likelihood estimator for $g$ under this constraint. We define the ``$n\sigma$'' confidence interval around the best fit $m_a$ by finding the interval over which,
\begin{equation}
2\ln \frac{\max{\mathcal{L}_1}}{\mathcal{L}_1(m_a)}<n^2 \, .
\end{equation}

Similar to the procedure for calculating our mass discovery limits, we can derive a `mass estimation limit', by finding the minimum coupling value that ensures that the mass be constrained to within a given precision. Formally, if the 2$\sigma$ confidence interval on $m_a$ is bounded from above and below by say $m_a(1-\epsilon_-)$ and $m_a(1+\epsilon_+)$, then the mass estimation limit for some accuracy $\epsilon$ is the minimum value of $g$ for which $\max(\epsilon_-,\epsilon_+)<\epsilon$.

In Fig.~\ref{fig:massestimation} we show a set of these mass estimation limits for several levels of precision, $\epsilon$. Here we show only the limits for the axion-photon coupling although the same result for $g_{ae}$ is similar. We can see that in general the shape of the curves follows the mass discovery limit (shown in orange) at low masses. In this regime the spectral topology stores very little information about the mass, and most of the reconstruction power is given essentially by small changes in the lowest energy part of the spectrum. The shape of the likelihood in this regime is smooth and well behaved. So the likelihood ratio for 2$\sigma$ away from the true mass and the the likelihood ratio relative to $m_a=0$ both scale in a similar way. This means that the mass estimation limits and the mass discovery limits look very similar. 

However we start to observe some differences at larger masses when the spectra have oscillations extending across the full range of energies. When $m_a \gtrsim 0.02$ the conversion spectrum picks up large peaks at particular energies. Since the energies of these peaks are finely controlled by the value of the mass, this gives rise to likelihoods with peculiar shapes where masses slightly off the true value are very highly disfavoured (as can be seen in the highest mass panel of Fig.~\ref{fig:Like}), even when the value of the coupling is too small to give enough statistics to fit the rest of the spectrum. At the very highest masses the spectral oscillations become so rapid that the small window of masses around the true value that were highly disfavoured get increasingly close together, until eventually they are lost to the energy resolution. Towards this regime, we see the mass estimation limits rise more steeply than the mass discovery limits. 

\subsection{Scaling results for other helioscopes}\label{sec:baby}
We have framed this discussion around the planned next generation helioscope IAXO which will be required to reach the QCD axion. However it is straightforward to use our results to consider other intermediate helioscopes that will be realised before IAXO. The scaling of the mass discovery limit, away from our result $g_{\rm IAXO}$, in terms of the relevant experimental parameters, is as follows:
\begin{align}
 g_{\rm disc}\left[m_a \left(\frac{L}{20 {\rm m}}\right)^{-\frac{1}{2}}\right] =& \, g_{\rm IAXO}(m_a) \, \nonumber \\
&\times \left( \frac{B}{2.5\, {\rm T}} \right)^{-\frac{1}{2}}
\left( \frac{S}{2.26\, {\rm m}^2} \right)^{-\frac{1}{4}} \nonumber \\
&\times \left( \frac{t}{3\, {\rm years}} \right)^{-\frac{1}{4}}  \left( \frac{L}{20\, {\rm m}} \right)^{-\frac{1}{2}} \nonumber \\
& \times \left( \frac{\varepsilon_{\rm D}\varepsilon_{\rm T}}{0.8\times 0.7} \right)^{-\frac{1}{4}} \, .
\end{align}
We express the discovery limit in this way to imply that a shorter $L$ shifts both the coupling and the mass to larger values. The cases for both Primakoff and axion-electron fluxes are identical, for $g$ corresponding to $g_{a\gamma}$ and $\sqrt{g_{ae}g_{a\gamma}}$ respectively.

To demonstrate this scaling we take the example of ``babyIAXO'', a smaller-scale helioscope planned as a test bed for the full IAXO. This experiment will use only one magnet bore rather than eight, reducing its $S$ by the same factor. It will also only have half the bore length $L=10$ m. For a total exposure time $t=1$ years the mass discovery limit of babyIAXO could reach a minimum coupling around a factor of 2.6 higher, whilst also shifted to higher masses by a factor of $\sim$1.4. We initially showed these limits in Fig.~\ref{fig:mainresult} as a dashed line inside the final IAXO result. Along similar lines, we can estimate the result for the medium-scale helioscope TASTE~\cite{Anastassopoulos:2017kag}. We use the projections of the latter reference which assume a magnetic field of 3.5 T with a single similarly sized bore with a length of 12 m. Assuming the same exposure as IAXO and a similar detector (i.e. the same $\varepsilon_{\rm D,T}$) we predict that TASTE can reach a mass discovery limit a factor of 1.42 times higher in coupling and a factor of 1.29 shifted upwards in mass.

\section{Conclusions}\label{sec:conc}
With an axion helioscope it is possible, with uniform sensitivity, to explore a wide range of masses below a certain critical value. However this leads to the problem that for this vast swath of low-mass axions and axion-like particles, the value of the mass is indistinguishable from zero. Whilst the injection of a buffer gas is an option to incrementally extend the sensitivity at relatively high masses, this method is not applicable for lower masses. We have demonstrated here for the first time that helioscopes do in fact have the capability of determining the axion mass in the vacuum mode (see Fig.~\ref{fig:mainresult} for this main result in the context of the full axion parameter space). 
We have found good prospects for the measurement of both solar axion fluxes (see Fig.~\ref{fig:fluxes}) generated via the axion-photon and axion-electron couplings. 

For axion masses below the buffer gas regime, but above the critical value at which the axion conversion spectrum is coherent across the magnet length, there is information about the mass is encoded in rapid oscillations in the x-ray spectrum (see Fig.~\ref{fig:spectra}). For smaller masses these oscillations occupy only the very lowest energy bins observable. So to realise the best results for IAXO---those that match what we have demonstrated here---good energy resolution and low detector energy thresholds will be required. Based on x-ray detection technologies currently under consideration~\cite{Irastorza:2013dav, Armengaud:2014gea}, this looks to be more than reasonable. We have demonstrated that $>3\sigma$ sensitivity to the axion mass is possible for axion masses down to $\sim 1-5 \times 10^{-3}$~eV (see Fig.~\ref{fig:result}). The highly characteristic spectral oscillations allow the value of the mass to not only be distinguished from zero, but also constrained to within percent-level accuracies (see Fig.~\ref{fig:massestimation}). 

At the upper end of the masses we have studied here, the suppressed number of expected events pushes the barrier on the mass measurement  to higher values of $g_{a\gamma}$ and $g_{ae}  g_{a\gamma}$, eventually rising above existing constraints at around $10^{-1}$ eV. Fortunately this is precisely the regime for which the buffer gas mode is ideal. In this mode the sensitivity of the helioscope is dramatically increased for a narrow range around the effective photon mass to which the chosen gas density corresponds. It is likely then that the mass sensitivity here will follow the discovery projections for IAXO, since the very detection of the axion in this mode requires that it has a mass. In follow-up work that may consider this regime in more detail, it will also be worthwhile to account for the theoretical systematic uncertainties on the axion flux.

A remaining question that we have not addressed here is whether IAXO can use this same spectral information to distinguish between the axion-photon and axion-electron coupling. It turns out that for many of the best-motivated axion models, including those which align with astrophysical hints~\cite{Giannotti:2017hny}, such model discrimination is indeed possible. Towards the completion of this work we were made aware of a study in preparation~\cite{Jaeckel:2018mbn}, which approached a similar type of analysis from this perspective. The results of this work deal with a more complete set of model parameters describing the axion, but use specific benchmark masses. On the other hand we have dealt with ranges of masses more generally, but have had to make an assumption about which coupling is dominantly controlling the axion signal. Hence the two studies are highly complementary in their approaches.

We have shown that over a limited range IAXO will have sensitivity to the masses of KSVZ axions. Some new theoretical predictions propose that this range of axion masses could be of importance in describing both inflation and dark matter~\cite{Visinelli:2009kt,Kawasaki:2014sqa,Ringwald:2015dsf,Daido:2017wwb,Co:2017mop,Gorghetto:2018myk}. Therefore IAXO, as well as being simply a search for solar axions, may additionally prove valuable in the search for dark matter, and a companion to future haloscopes searching in the same range~\cite{Baryakhtar:2018doz}. Alternatively if the axion-like particle does not possess a measurable mass then this could help constrain any further astrophysical searches.

\acknowledgments
The authors acknowledge the resources from the supercomputer Cierzo, the HPC infrastructure of the Centro de Supercomputaci\'on de Arag\'on (CESAR) and technical expertise and assistance provided by BIFI (Universidad de Zaragoza). We acknowledge the support from the Spanish Ministry of Economy and Competitiveness (MINECO) under Grants No. FPA2013-41085-P and No. FPA2016-76978-C3-1-P. TD acknowledges the support from the Ram\'on y Cajal program of MINECO. The authors also acknowledge the support from the Croatian Science Foundation under the Project No. IP-2014-09-3720. CAJO is supported by the grant FPA2015-65745-P (MINECO/FEDER). IGI acknowledges support from the European Research Council (ERC) under the European Union's Horizon 2020 research and innovation programme  (Grant No. 788781 / ERC-AdG IAXO+). JR is supported by the Ramon y Cajal Fellowship 2012-10597, the Grant No. FPA2015-65745-P (MINECO/FEDER), the EU through the ITN ``Elusives'' H2020-MSCA-ITN-2015/674896 and the Deutsche Forschungsgemeinschaft under Grant No. SFB-1258 as a Mercator Fellow. 

\bibliographystyle{apsrev4-1}
\bibliography{ref}

\end{document}